\documentclass{Interspeech}
\usepackage{times}
\usepackage{latexsym}
\usepackage{booktabs}
\usepackage{verbatim}
\usepackage{multirow}
\usepackage{colortbl}  
\usepackage[table,xcdraw]{xcolor}
\usepackage{booktabs}
\usepackage[T1]{fontenc}
\usepackage[utf8]{inputenc}
\usepackage{microtype}
\usepackage{inconsolata}
\usepackage{adjustbox}
\usepackage{tikz} 
\usepackage{xcolor} 
\usepackage{ subcaption, algorithm2e, url, float, etoolbox, adjustbox, pgf,booktabs, verbatim}
\usepackage{xcolor}
\usepackage{amsmath}
\usepackage{amssymb}
\usepackage{tcolorbox}
\renewcommand{\thefootnote}{\fnsymbol{footnote}}

\usepackage{etoolbox}
\makeatletter
\patchcmd{\affiliation}
  {\renewcommand{\affiliationsep}{\vskip1pt}}
  {\renewcommand{\affiliationsep}{, }}
  {}{}
\makeatother
\definecolor{litepurple5}{RGB}{237,231,246}   
\definecolor{litepurple10}{RGB}{209,196,233}
\definecolor{litepurple15}{RGB}{179,157,219}
\definecolor{litepurple20}{RGB}{149,117,205}
\definecolor{litepurple25}{RGB}{126,87,194}    

\title{\textbf{\texttt{PARROT}}: Synergizing Mamba and Attention-based SSL Pre-Trained Models via Parallel Branch Hadamard Optimal Transport for Speech Emotion Recognition}

\author[affiliation={1}]{Orchid Chetia}{Phukan*}
\author[affiliation={1,2}]{Mohd Mujtaba}{Akhtar*} 
\author[affiliation={1,3}]{Girish*}{}
\author[affiliation={4}]{Swarup Ranjan}{Behera}
\author[affiliation={4}]{Jaya Sai Kiran}{Patibandla}
\author[affiliation={1}]{Arun Balaji}{Buduru}
\author[affiliation={5,6}]{Rajesh}{Sharma}
\affiliation{}{IIIT-Delhi}{India}
\affiliation{}{V.B.S.P.U}{India}
\affiliation{}{UPES}{India}
\affiliation{}{Independent Researcher}{India}
\affiliation{}{University of Tartu}{Estonia}
\affiliation{}{Plaksha University}{India}
\email{\textcolor{blue}{\texttt{Correspondence:}} orchidp@iiitd.ac.in} 

\usepackage{comment}
\begin{document}

\maketitle
\begingroup
  \renewcommand{\thefootnote}{\fnsymbol{footnote}}
  \setcounter{footnote}{0}
  \footnotetext{* Contributed equally as a first authors.}
\endgroup

\begin{abstract}
The emergence of Mamba as an alternative to attention-based architectures has led to the development of Mamba-based self-supervised learning (SSL) pre-trained models (PTMs) for speech and audio processing. Recent studies suggest that these models achieve comparable or superior performance to state-of-the-art (SOTA) attention-based PTMs for speech emotion recognition (SER). 
Motivated by prior work demonstrating the benefits of PTM fusion across different speech processing tasks, we hypothesize that leveraging the complementary strengths of Mamba-based and attention-based PTMs will enhance SER performance beyond the fusion of homogenous attention-based PTMs. To this end, we introduce a novel framework, \textbf{\texttt{PARROT}} that integrates parallel branch fusion with Optimal Transport and Hadamard Product. Our approach achieves SOTA results against individual PTMs, homogeneous PTMs fusion, and baseline fusion techniques, thus, highlighting the potential of heterogeneous PTM fusion for SER. 

\end{abstract}

\section{Introduction}
Speech Emotion Recognition (SER) bridges human-computer interaction, finds applications in mental health monitoring as well as in empathetic AI systems~\cite{Tahon2016TowardsAS,Souganciouglu2020IsEF}. It enables machines to understand and respond to human emotions, fostering more natural and intuitive interactions. Traditional SER research often employs handcrafted features such as MFCCs, which capture the spectral properties of speech and have proven effective in representing emotional cues. These features were initially modelled with classical ML techniques such as SVM \cite{dahake2016speaker}, tree-based methods \cite{noroozi2017vocal}. This was followed by the use of deep learning techniques \cite{abbaschian2021deep, zhao2019speech}. \newline
By the end of end of last decade, the landscape of SER research has changed for the better through the use and the wide scale availability of self-supervised learning (SSL) pre-trained models (PTMs). These PTMs trained on large-scale diverse data provides performance benefits as well as discard the necessity of training models from scratch. These PTMs have led to significant development in SER. As such, researchers have explored various state-of-the-art (SOTA) PTMs \cite{10066578, atmaja2022evaluating, osman24_interspeech, phukan24b_interspeech}. Pepino et al. \cite{pepino21_interspeech} used wav2vec2 with LSTM and CNN downstreams and showed its effectiveness in comparison to conventational features such as eGeMAPS and spectrogram. Morais et al. \cite{morais2022speech} gave a comprehensive comparison of various SSL PTMs such as wav2vec2, HuBERT with different downstream networks. These PTMs have predominantly utilized attention-based architectures, which excel at capturing contextual dependencies in speech signals. \newline 
In recent times , an alternative architecture type of PTMs has captured attention in the community: mamba-based PTMs \cite{yadav24_interspeech, 10720871, shams2024ssamba}. These mamba-based PTMs are based on top of mamba architecture which is structured state-space model (SSM) and has emerged as a promising alternative due to its ability to efficiently model long-range dependencies with linear complexity \cite{gu2023mamba}. These mamba-based PTMs have set benchmarks in sequence modeling tasks and initial research suggest that mamba-based PTMs perform competitively with, or even surpass, attention-based PTMs in SER \cite{yadav24_interspeech}. Prior research has also shown the benefits of PTM fusion for SER \cite{wu2023investigation} due to their complementary behavior. This is also observed across various speech processing tasks such as speech recognition \cite{arunkumar22b_interspeech} and speech deepfake detection \cite{chetia-phukan-etal-2024-heterogeneity}. However, previous works have integrated only attention-based models. \newline
In this study, we fuse mamba and attention-based PTMs for SER and \textit{hypothesize that the fusion of these heterogeneous PTMs will yield richer and more robust representations for improved SER as attention-based PTMs will capture intricate global dependencies while mamba-based PTMs excels at efficient long-range processing}. We are the first study to the best of our knowledge to explore fusion of such heterogeneous PTMs for SER. To our end, we propose, \textbf{\texttt{PARROT}} (\texttt{\textbf{PAR}}allel B\textbf{\texttt{R}}anch Hadamard \textbf{\texttt{O}}ptimal \textbf{\texttt{T}}ransport), a novel framework to align and integrate heterogeneous mamba and attention-based PTMs. It employs parallel branch fusion, incorporating the hadamard product and optimal transport. Hadamard product captures local interactions by performing element-wise operations between the representations, preserving fine-grained details. Meanwhile, Optimal Transport operates at a global scale, aligning the distributions of features across the two PTMs, ensuring that the fused representations are coherent and effectively integrated for improved performance. \newline
\noindent The key contributions of our study are as follows:
\begin{itemize}
    \item We introduce \textbf{\texttt{PARROT}}, a novel framework that encompasses parallel branch fusion with Hadamard Product and Optimal Transport that inherently captures local interactions and global interaction for improved SER performance.
    \item With \textbf{\texttt{PARROT}} through the synergy of mamba and attention-based PTMs, we achieve the topmost most performance across different SER datasets (CREMA-D (\textit{English}), emo-DB (\textit{German}), MESD (\textit{Mexican Spanish})) than individual PTMs and homogeneous fusion of attention-based PTMs. It also reports better performance in comparison to baseline fusion methods. These PTMs are SOTA SSL PTMs for SER, thus, \textbf{\texttt{PARROT}} achieves SOTA performance for SER with its heterogeneous fusion.
\end{itemize}
\noindent All code and models used in this study are accessible at: \url{https://github.com/Helix-IIIT-Delhi/PARROT-SER}

\begin{figure*}[hbt!]
    \centering
    \includegraphics[width=0.55\textwidth]{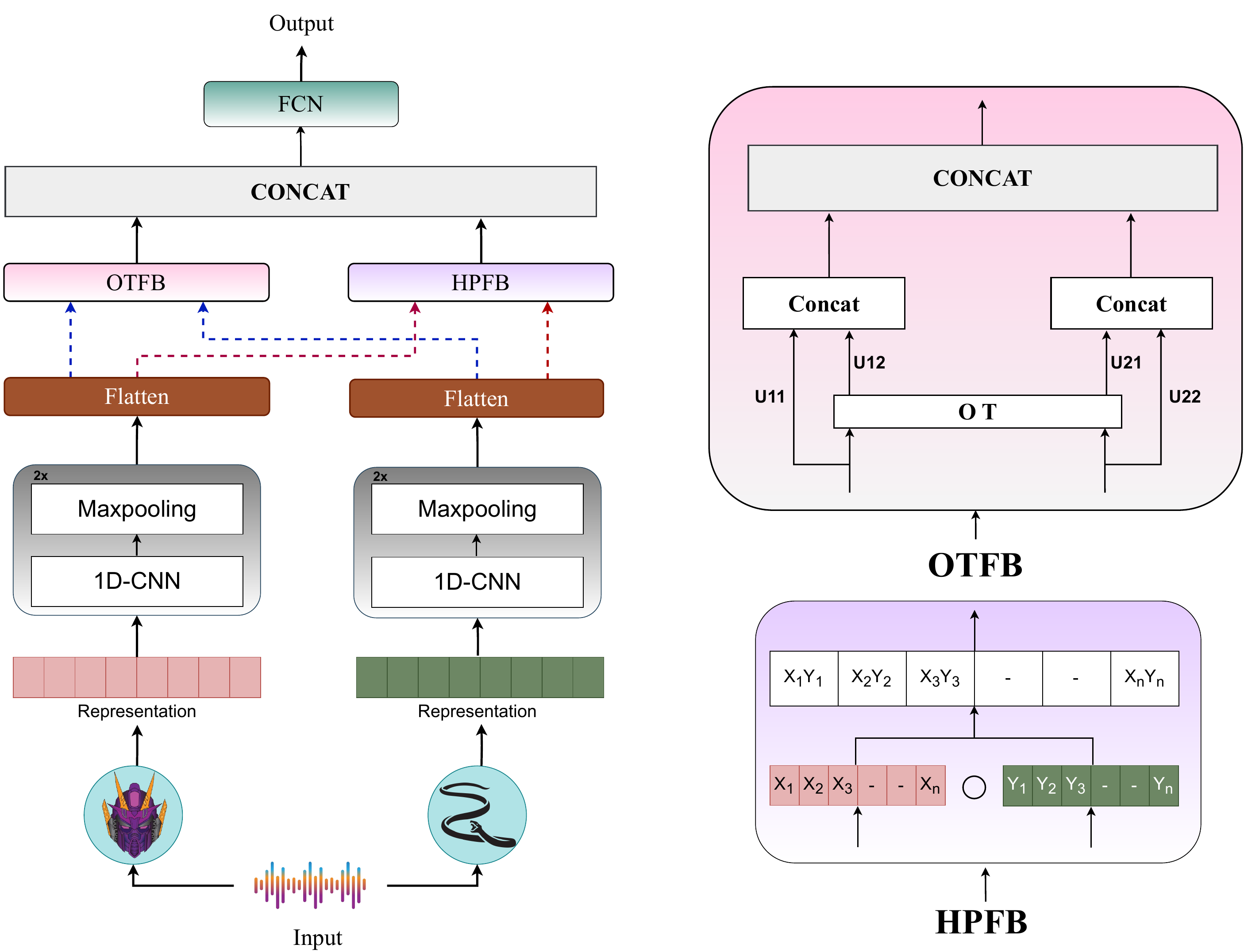}
    \caption{Proposed Framework: \textbf{\texttt{PARROT}}; OTFB, HPFB stands for Optimal Transport Fusion Block and Hadamard Product Fusion Block respectively; $U_11$, $U_22$ represents the representational space of PTM 1 and PTM2; $U_12$, $U_21$ represents the transported space of PTM 2 to PTM 1 and vice versa}
    \label{fig:parrot_framework}
\end{figure*}

\section{Pre-Trained Models}

\noindent In this section, we discuss the PTMs considered in our study.

\noindent \textbf{Audio-MAMBA} \cite{yadav24_interspeech}\footnote{\url{https://github.com/SarthakYadav/audio-mamba-official?tab=readme-ov-file}}: Audio Mamba is a selective state space model that is trained in a self-supervised fusion to learn general-purpose representations from randomly masked spectrogram patches. Trained on the AudioSet dataset, it outperforms its attention-based counterparts baselines across diverse speech and audio tasks including SER. We use the tiny, small, and base versions of 4.8M, 17.9M, and 69.3M parameters.\par 

\noindent \textbf{WavLM} \cite{chen2022wavlm}\footnote{\url{https://huggingface.co/microsoft/wavlm-base}}: It is a SOTA attention-based PTM on SUPERB that integrates masked speech modeling with denoising objectives during its pre-training, effectively learning robust representations from noisy and clean speech alike. We have used the base version with 94.70M parameters and trained on librispeech 960 hours of english speech.  \par

\noindent \textbf{UniSpeech-SAT} \cite{chen2022unispeech}\footnote{\url{https://huggingface.co/microsoft/unispeech-sat-base}}: It is also a SOTA attention-based PTM on SUPERB and trained in a self-supervised fashion with speaker-aware multi-task learning. We utilize the base version of 94.68M parameters and pre-trained on 960 hours of librispeech english data. \par

\noindent \textbf{Wav2vec2} \cite{baevski2020wav2vec}\footnote{\url{https://huggingface.co/facebook/wav2vec2-base}}: This contrastive SSL attention-based PTM that learns speech representations by masking segments of latent features. We use its base version trained on librispeech 960 hours english data with 95.04M parameters. Wav2vec2 improves previous SOTA methods in speech recognition.    \par

\noindent \textbf{HuBERT} \cite{hsu2021hubert}\footnote{\url{https://huggingface.co/facebook/hubert-base-ls960}}: HuBERT employs a SSL framework that iteratively refines its representations using k-means clustering and solves a BERT-like masked prediction objective. HuBERT improves over Wav2vec2 in speech recognition. We use the base version trained on english 960 hours librispeech data with 94.68M parameters. \par

\noindent \textbf{Massively Multilingual Speech (MMS)} \cite{pratap2024scaling}\footnote{\url{https://huggingface.co/facebook/mms-1b}}: It is attention-based SSL PTM built on top of Wav2vec2 architecture. It extends pre-training to almost 1400 languages. It improves over XLS-R and Whisper in various multilingual speech processing. We use the 1B parameters version in our experiments. \par

\noindent We extract representations from the last hidden state of the frozen mamba and attention-based PTMs by pooling average. We representations are of dimensions: 768 for WavLM, Unispeech-SAT, Wav2vec2 and HuBERT; 1280 for MMS; 960 for tiny, 1920 for small, 3840 for base versions of Audio-MAMBA. \par

\section{Modeling Pipeline}
In this section, we discuss the downstream modeling networks to be employed with individual PTMs and the proposed framework for aligning PTMs, \textbf{\texttt{PARROT}}. We make use of SVM, Fully Connected Network (FCN), and CNN as the downstreams modeling with individual PTMs. For SVM, we kept the default hyperparameters. For CNN, we make use of two 1D convolutional layers with 64 and 128 filters, respectively, and a kernel size of 3 with ReLU activation function. After each 1D convolutional layer we attach a maxpooling. The output is then flattened and passed through a FCN containing a dense layer with 128 neurons and ReLU activation, followed by a softmax layer for multi-class classification. For FCN, we keep the modeling same as FCN used in CNN downstream. \par 


\subsection{PARROT}

We propose \texttt{\textbf{PARROT}}, a novel framework for integrating PTMs. The architecture of the proposed framework is presented in Figure \ref{fig:parrot_framework}. First, the extracted representations from both PTMs are processed through two 1D convolutional blocks with the same number of filters as used in Individual representation modeling above. Also, the rest modeling remains same. After flattening, the outputs are linearly projected into a 120-dimensional latent space, ensuring computational efficiency while retaining expressive information. Then, \texttt{\textbf{PARROT}} employs a parallel branch fusion strategy that combines local feature interactions via Hadamard product (HP) and global distribution alignment via Optimal Transport (OT). HP performs element-wise multiplication between the representations \( \mathbf{R}_p \)  and \( \mathbf{R}_q \) of the PTMs and is given by  
$\mathbf{HP} = \mathbf{R}_p \odot \mathbf{R}_q$, where \( \odot \) denotes element-wise multiplication. While HP fusion retains structural details, it does not ensure global coherence between the PTMs. To address this, we employ OT to align the distributions of feature representations from PTMs. First, we compute the cost matrix \( C \) using the normalized Euclidean distance between the PTMs feature matrices \( \mathbf{R}_p \) and \( \mathbf{R}_q \) :

\begin{align}
  C &= \frac{\left|| R_p - R_q \right||_2}
  {\max (\left|| R_p - R_q \right||_2)}
  \label{equation:eq1}
\end{align}

\noindent To efficiently compute the transport plan, we apply the Sinkhorn algorithm. The OT plan \( \Gamma \) is then derived as: $\Gamma = \text{Sinkhorn}(C)$. Using \( \Gamma \), we transport features between PTMs to enforce distributional alignment, mapping \( \mathbf{R}_p \) into \( \mathbf{R}_q \)'s space and vice versa: $ \mathbf{R}_p \to \mathbf{R}_q = \Gamma \cdot \mathbf{R}_p, \quad \mathbf{R}_q \to \mathbf{R}_p = \Gamma^T \cdot \mathbf{R}_q $. These transported features are then concatenated with the original representations to form the final fused representations $
\mathbf{F} = \text{Concat}(\mathbf{F}_q, \mathbf{F}_p)$: 
$\mathbf{F}_q = \text{Concat}(\mathbf{R}_P \to \mathbf{R}_q, \mathbf{R}_q) $, $\mathbf{F}_p = \text{Concat}(\mathbf{R}_q \to \mathbf{R}_q, \mathbf{R}_p) $. Finally, the fused features from both branches are concatenated and passed through a FCN with a dense layer of 128 neurons with softmax activation function for the final prediction. By preserving both local feature interactions and enforcing global alignment, \texttt{\textbf{PARROT}} enables robust fusion of PTMs. The trainable parameters of \texttt{\textbf{PARROT}} vary between 3.2M and 13M.


\section{Experiments}

\subsection{Benchmark Datasets}

\noindent \textbf{Crowd-Sourced Emotional Multimodal Actors Dataset (CREMA-D) \cite{cao2014crema}} serves as a widely recognized benchmark for SER and comprising 7,442 utterances from 48 male and 43 female actors, it spans a diverse range of speaker ages and ethnicities. This dataset includes six distinct emotional categories: anger, happiness, sadness, fear, disgust, and neutral, with each actor contributing 12 unique sentences. \textbf{German Emotional Speech Database (Emo-DB) \cite{burkhardt2005database}} is German benchmark SER database and containing 535 utterances from ten actors (five male and five female). The dataset features seven emotional states: anger, anxiety/fear, boredom, disgust, happiness, neutral, and sadness. \textbf{The Mexican Emotional Speech Database (MESD) \cite{duville2021mexican}} is Mexican-spanish database containing 864 utterances representing six emotional states: anger, disgust, fear, happiness, neutral, and sadness. \newline
\noindent\textbf{Training Details}: All models are trained using the Adam optimizer with cross-entropy loss. The learning rate is 1e-3, batch size 32, and training runs for 50 epochs. To mitigate overfitting, we use dropout and early stopping. We follow five-fold cross-validation, with four folds for training and one for testing.

\begin{table}[hbt!]
\scriptsize
\setlength{\tabcolsep}{7pt}
\centering
\begin{tabular}{l|ll|ll|ll}
\toprule
\multicolumn{1}{c|}{\textbf{PTM's}} & \multicolumn{2}{c}{\textbf{CREMA-D}} & \multicolumn{2}{c}{\textbf{Emo-DB}} & \multicolumn{2}{c}{\textbf{MESD}} \\ 
\midrule
                          & \textbf{Acc} & \textbf{F1} & \textbf{Acc} & \textbf{F1} & \textbf{Acc} & \textbf{F1} \\
\midrule
\multicolumn{7}{c}{\textbf{SVM}} \\
\cmidrule{1-7}
A(T) & \cellcolor{litepurple5}60.96 & \cellcolor{litepurple5}59.85 & \cellcolor{litepurple5}68.96 & \cellcolor{litepurple5}68.10 & \cellcolor{litepurple10}70.96 & \cellcolor{litepurple10}69.63 \\
A(S) & \cellcolor{litepurple15}68.96 & \cellcolor{litepurple15}67.10 & \cellcolor{litepurple5}78.96 & \cellcolor{litepurple5}77.65 & \cellcolor{litepurple10}71.63 & \cellcolor{litepurple10}70.85 \\
A(B) & \cellcolor{litepurple10}66.99 & \cellcolor{litepurple10}65.25 & \cellcolor{litepurple5}81.65 & \cellcolor{litepurple5}80.94 & \cellcolor{litepurple10}75.69 & \cellcolor{litepurple10}74.62 \\
W    & \cellcolor{litepurple5}61.92  & \cellcolor{litepurple5}60.36 & \cellcolor{litepurple10}85.09 & \cellcolor{litepurple10}84.63 & \cellcolor{litepurple5}45.96  & \cellcolor{litepurple5}44.12 \\
H    & \cellcolor{litepurple10}64.25 & \cellcolor{litepurple10}63.98 & \cellcolor{litepurple10}86.16 & \cellcolor{litepurple10}85.31 & \cellcolor{litepurple5}60.94  & \cellcolor{litepurple5}59.76 \\
W2   & \cellcolor{litepurple5}59.86  & \cellcolor{litepurple5}58.23 & \cellcolor{litepurple15}86.31 & \cellcolor{litepurple15}85.93 & \cellcolor{litepurple5}61.37  & \cellcolor{litepurple5}60.88 \\
U    & \cellcolor{litepurple5}62.93  & \cellcolor{litepurple5}61.09 & \cellcolor{litepurple5}77.69  & \cellcolor{litepurple5}76.38 & \cellcolor{litepurple5}33.95  & \cellcolor{litepurple5}32.16 \\
M    & \cellcolor{litepurple10}66.94 & \cellcolor{litepurple10}65.28 & \cellcolor{litepurple5}70.11  & \cellcolor{litepurple5}69.89 & \cellcolor{litepurple15}81.03 & \cellcolor{litepurple15}80.07 \\
\midrule
\multicolumn{7}{c}{\textbf{FCN}} \\
\cmidrule{1-7}
A(T) & \cellcolor{litepurple5}62.96 & \cellcolor{litepurple5}61.85 & \cellcolor{litepurple5}70.96 & \cellcolor{litepurple5}69.36 & \cellcolor{litepurple5}72.96 & \cellcolor{litepurple5}71.85 \\
A(S) & \cellcolor{litepurple15}69.52 & \cellcolor{litepurple15}68.20 & \cellcolor{litepurple10}80.66 & \cellcolor{litepurple10}79.63 & \cellcolor{litepurple10}76.93 & \cellcolor{litepurple10}75.11 \\
A(B) & \cellcolor{litepurple10}68.41 & \cellcolor{litepurple10}67.11 & \cellcolor{litepurple15}83.63 & \cellcolor{litepurple15}82.89 & \cellcolor{litepurple15}77.92 & \cellcolor{litepurple15}76.13 \\
W    & \cellcolor{litepurple5}62.96  & \cellcolor{litepurple5}61.08 & \cellcolor{litepurple15}87.85 & \cellcolor{litepurple15}86.36 & \cellcolor{litepurple5}47.98  & \cellcolor{litepurple5}46.21 \\
H    & \cellcolor{litepurple10}66.29 & \cellcolor{litepurple10}65.85 & \cellcolor{litepurple15}87.33 & \cellcolor{litepurple15}86.21 & \cellcolor{litepurple5}61.20  & \cellcolor{litepurple5}60.96 \\
W2   & \cellcolor{litepurple5}61.01  & \cellcolor{litepurple5}60.36 & \cellcolor{litepurple15}88.63 & \cellcolor{litepurple15}87.03 & \cellcolor{litepurple5}62.78  & \cellcolor{litepurple5}61.96 \\
U    & \cellcolor{litepurple5}64.82  & \cellcolor{litepurple5}63.33 & \cellcolor{litepurple5}79.65  & \cellcolor{litepurple5}78.51 & \cellcolor{litepurple5}36.96  & \cellcolor{litepurple5}35.21 \\
M    & \cellcolor{litepurple10}67.52 & \cellcolor{litepurple10}66.36 & \cellcolor{litepurple5}71.39  & \cellcolor{litepurple5}70.88 & \cellcolor{litepurple15}82.66 & \cellcolor{litepurple15}81.39 \\
\midrule
\multicolumn{7}{c}{\textbf{CNN}} \\
\cmidrule{1-7}
A(T) & \cellcolor{litepurple5}63.60 & \cellcolor{litepurple5}63.60 & \cellcolor{litepurple5}71.03 & \cellcolor{litepurple5}70.14 & \cellcolor{litepurple5}73.99 & \cellcolor{litepurple5}73.93 \\
A(S) & \cellcolor{litepurple10}69.51 & \cellcolor{litepurple10}69.49 & \cellcolor{litepurple5}81.31 & \cellcolor{litepurple5}79.27 & \cellcolor{litepurple5}78.61 & \cellcolor{litepurple5}78.53 \\
A(B) & \cellcolor{litepurple10}69.91 & \cellcolor{litepurple15}69.90 & \cellcolor{litepurple5}84.11 & \cellcolor{litepurple5}82.90 & \cellcolor{litepurple5}78.03 & \cellcolor{litepurple5}77.96 \\
W    & \cellcolor{litepurple5}67.96  & \cellcolor{litepurple5}66.25 & \cellcolor{litepurple15}89.14 & \cellcolor{litepurple15}88.69 & \cellcolor{litepurple5}48.55  & \cellcolor{litepurple5}48.16 \\
H    & \cellcolor{litepurple15}69.95 & \cellcolor{litepurple10}68.23 & \cellcolor{litepurple10}88.26 & \cellcolor{litepurple10}87.11 & \cellcolor{litepurple5}62.43  & \cellcolor{litepurple5}62.43 \\
W2   & \cellcolor{litepurple5}61.18  & \cellcolor{litepurple5}60.88 & \cellcolor{litepurple15}90.01 & \cellcolor{litepurple15}89.62 & \cellcolor{litepurple5}63.01  & \cellcolor{litepurple5}63.24 \\
U    & \cellcolor{litepurple5}65.28  & \cellcolor{litepurple5}64.11 & \cellcolor{litepurple5}81.36 & \cellcolor{litepurple5}80.96 & \cellcolor{litepurple5}38.15 & \cellcolor{litepurple5}37.91 \\
M    & \cellcolor{litepurple10}68.25 & \cellcolor{litepurple5}67.96 & \cellcolor{litepurple5}72.90 & \cellcolor{litepurple5}66.24 & \cellcolor{litepurple15}83.24 & \cellcolor{litepurple15}83.10 \\
\bottomrule
\end{tabular}
\caption{Evaluation Scores; Scores are in \% and average of five-folds: Abbreviations used are: Audio-mamba (Tiny A(T), Small A(S), Base A(B)), WavLM (W), HuBERT (H), Wav2vec2 (W2), Unispeech-SAT (U), and MMS (M); Acc, F1 stands for Accuracy and macro average F1 score; Abbreviations used in this Table \ref{single} are kept same for Table \ref{fusion}}
\label{single}
\end{table}

\begin{table*}[hbt!]
\scriptsize
\setlength{\tabcolsep}{8pt}
\centering
\begin{tabular}{l|ll|ll|ll|ll|ll|ll}
\toprule
\multicolumn{1}{c|}{\textbf{}} & \multicolumn{4}{c|}{\textbf{CREMA-D}} & \multicolumn{4}{c|}{\textbf{Emo-DB}} & \multicolumn{4}{c}{\textbf{MESD}} \\ 
\midrule
                                    & \multicolumn{2}{c|}{\textbf{Concatenation}} & \multicolumn{2}{c|}{\textbf{\texttt{PARROT}}} & \multicolumn{2}{c|}{\textbf{Concatenation}} & \multicolumn{2}{c|}{\textbf{\texttt{PARROT}}} & \multicolumn{2}{c|}{\textbf{Concatenation}} & \multicolumn{2}{c}{\textbf{\texttt{PARROT}}} \\ 
\cmidrule(lr){2-5} \cmidrule(lr){6-9} \cmidrule(lr){10-13}
\multicolumn{1}{l|}{\textbf{Fusion}}               & \textbf{Acc} & \textbf{F1} & \textbf{Acc} & \textbf{F1} & \textbf{Acc} & \textbf{F1} & \textbf{Acc} & \textbf{F1} & \textbf{Acc} & \textbf{F1} & \textbf{Acc} & \textbf{F1} \\ 
\midrule
A(T)+W   & \cellcolor{litepurple10}63.37 & \cellcolor{litepurple10}62.89 & \cellcolor{litepurple10}64.38 & \cellcolor{litepurple10}63.38 & \cellcolor{litepurple15}76.96 & \cellcolor{litepurple15}75.61 & \cellcolor{litepurple15}78.69 & \cellcolor{litepurple15}77.25 & \cellcolor{litepurple5}44.38 & \cellcolor{litepurple5}43.81 & \cellcolor{litepurple5}46.85 & \cellcolor{litepurple5}45.28 \\
A(T)+H   & \cellcolor{litepurple15}65.54 & \cellcolor{litepurple15}64.86 & \cellcolor{litepurple15}66.82 & \cellcolor{litepurple15}65.76 & \cellcolor{litepurple10}76.28 & \cellcolor{litepurple10}75.58 & \cellcolor{litepurple10}77.62 & \cellcolor{litepurple10}76.28 & \cellcolor{litepurple5}44.85 & \cellcolor{litepurple5}43.38 & \cellcolor{litepurple5}45.28 & \cellcolor{litepurple5}44.39 \\
A(T)+W2  & \cellcolor{litepurple5}60.03  & \cellcolor{litepurple5}59.94  & \cellcolor{litepurple5}61.76  & \cellcolor{litepurple5}60.58  & \cellcolor{litepurple5}75.14  & \cellcolor{litepurple5}74.61  & \cellcolor{litepurple5}76.93  & \cellcolor{litepurple5}75.58  & \cellcolor{litepurple5}45.03  & \cellcolor{litepurple5}44.97  & \cellcolor{litepurple5}46.92  & \cellcolor{litepurple5}45.22 \\
A(T)+U   & \cellcolor{litepurple5}62.56  & \cellcolor{litepurple5}61.14  & \cellcolor{litepurple5}63.94  & \cellcolor{litepurple5}62.28  & \cellcolor{litepurple5}74.94  & \cellcolor{litepurple5}73.39  & \cellcolor{litepurple5}75.58  & \cellcolor{litepurple5}75.16  & \cellcolor{litepurple5}44.34  & \cellcolor{litepurple5}43.08  & \cellcolor{litepurple5}45.34  & \cellcolor{litepurple5}44.28 \\
A(T)+M   & \cellcolor{litepurple5}62.95  & \cellcolor{litepurple5}61.59  & \cellcolor{litepurple5}63.86  & \cellcolor{litepurple5}62.47  & \cellcolor{litepurple15}76.85 & \cellcolor{litepurple15}75.28 & \cellcolor{litepurple15}77.78 & \cellcolor{litepurple15}76.94 & \cellcolor{litepurple5}45.37  & \cellcolor{litepurple5}44.65  & \cellcolor{litepurple5}46.39  & \cellcolor{litepurple5}45.80 \\
A(S)+W   & \cellcolor{litepurple10}63.48 & \cellcolor{litepurple10}62.94 & \cellcolor{litepurple10}64.45 & \cellcolor{litepurple10}63.94 & \cellcolor{litepurple15}77.26 & \cellcolor{litepurple15}76.18 & \cellcolor{litepurple15}79.54 & \cellcolor{litepurple15}78.51 & \cellcolor{litepurple5}46.88  & \cellcolor{litepurple5}45.82  & \cellcolor{litepurple5}47.58  & \cellcolor{litepurple5}46.64 \\
A(S)+H   & \cellcolor{litepurple5}62.14  & \cellcolor{litepurple5}61.19  & \cellcolor{litepurple5}63.68  & \cellcolor{litepurple5}62.24  & \cellcolor{litepurple15}77.36 & \cellcolor{litepurple15}76.52 & \cellcolor{litepurple15}78.82 & \cellcolor{litepurple15}77.34 & \cellcolor{litepurple5}45.28  & \cellcolor{litepurple5}44.28  & \cellcolor{litepurple5}46.62  & \cellcolor{litepurple5}45.82 \\
A(S)+W2  & \cellcolor{litepurple5}61.94  & \cellcolor{litepurple5}60.68  & \cellcolor{litepurple5}62.34  & \cellcolor{litepurple5}61.58  & \cellcolor{litepurple5}76.95 & \cellcolor{litepurple5}75.28 & \cellcolor{litepurple5}77.98 & \cellcolor{litepurple5}76.25 & \cellcolor{litepurple5}46.82  & \cellcolor{litepurple5}45.25  & \cellcolor{litepurple5}47.68  & \cellcolor{litepurple5}46.62 \\
A(S)+U   & \cellcolor{litepurple5}61.64  & \cellcolor{litepurple5}60.17  & \cellcolor{litepurple5}62.84  & \cellcolor{litepurple5}61.34  & \cellcolor{litepurple5}75.82 & \cellcolor{litepurple5}74.36 & \cellcolor{litepurple5}76.52 & \cellcolor{litepurple5}75.94 & \cellcolor{litepurple5}44.29  & \cellcolor{litepurple5}43.57  & \cellcolor{litepurple5}46.94  & \cellcolor{litepurple5}45.82 \\
A(S)+M   & \cellcolor{litepurple10}63.13 & \cellcolor{litepurple10}62.29 & \cellcolor{litepurple10}64.86 & \cellcolor{litepurple10}63.34 & \cellcolor{litepurple5}74.22  & \cellcolor{litepurple5}73.58  & \cellcolor{litepurple5}75.36  & \cellcolor{litepurple5}74.15  & \cellcolor{litepurple5}46.82  & \cellcolor{litepurple5}45.28  & \cellcolor{litepurple15}48.96 & \cellcolor{litepurple15}47.28 \\
A(B)+W   & \cellcolor{litepurple10}63.96 & \cellcolor{litepurple10}62.10 & \cellcolor{litepurple15}67.56 & \cellcolor{litepurple15}67.44 & \cellcolor{litepurple15}79.41 & \cellcolor{litepurple15}78.64 & \cellcolor{litepurple15}80.37 & \cellcolor{litepurple15}78.56 & \cellcolor{litepurple5}47.31  & \cellcolor{litepurple5}46.85  & \cellcolor{litepurple5}49.13  & \cellcolor{litepurple5}48.87 \\
\textbf{A(B)+H}   & \cellcolor{litepurple15}\textbf{70.63} & \cellcolor{litepurple15}\textbf{69.79} & \cellcolor{litepurple15}\textbf{73.68} & \cellcolor{litepurple15}\textbf{72.90} & \cellcolor{litepurple15}\textbf{89.92} & \cellcolor{litepurple15}\textbf{88.24} & \cellcolor{litepurple15}\textbf{92.24} & \cellcolor{litepurple15}\textbf{91.53} & \cellcolor{litepurple10}58.31  & \cellcolor{litepurple10}56.08  & \cellcolor{litepurple10}59.54  & \cellcolor{litepurple10}58.94 \\
A(B)+W2  & \cellcolor{litepurple10}69.96 & \cellcolor{litepurple10}69.21 & \cellcolor{litepurple10}71.98 & \cellcolor{litepurple10}70.94 & \cellcolor{litepurple10}88.94 & \cellcolor{litepurple10}87.64 & \cellcolor{litepurple10}89.44 & \cellcolor{litepurple10}88.57 & \cellcolor{litepurple10}56.37 & \cellcolor{litepurple10}55.39 & \cellcolor{litepurple10}57.23 & \cellcolor{litepurple10}56.64 \\
A(B)+U   & \cellcolor{litepurple5}61.27  & \cellcolor{litepurple5}60.96  & \cellcolor{litepurple5}62.53  & \cellcolor{litepurple5}62.75  & \cellcolor{litepurple5}73.46  & \cellcolor{litepurple5}72.68  & \cellcolor{litepurple5}74.77  & \cellcolor{litepurple5}73.17  & \cellcolor{litepurple5}37.91  & \cellcolor{litepurple5}36.49  & \cellcolor{litepurple5}38.15  & \cellcolor{litepurple5}36.77 \\
A(B)+M   & \cellcolor{litepurple5}61.25  & \cellcolor{litepurple5}60.97  & \cellcolor{litepurple5}63.26  & \cellcolor{litepurple5}63.24  & \cellcolor{litepurple5}65.96  & \cellcolor{litepurple5}64.05  & \cellcolor{litepurple5}66.36  & \cellcolor{litepurple5}57.29  & \cellcolor{litepurple15}\textbf{66.37} & \cellcolor{litepurple15}\textbf{65.28} & \cellcolor{litepurple15}\textbf{69.05} & \cellcolor{litepurple15}\textbf{68.72} \\
W+H      & \cellcolor{litepurple10}68.99 & \cellcolor{litepurple10}67.64 & \cellcolor{litepurple10}69.91 & \cellcolor{litepurple10}69.85 & \cellcolor{litepurple15}80.96 & \cellcolor{litepurple15}79.64 & \cellcolor{litepurple15}81.31 & \cellcolor{litepurple15}79.90 & \cellcolor{litepurple5}53.94  & \cellcolor{litepurple5}52.17  & \cellcolor{litepurple5}54.34  & \cellcolor{litepurple5}54.12 \\
W+W2     & \cellcolor{litepurple10}67.96 & \cellcolor{litepurple10}66.21 & \cellcolor{litepurple10}68.30 & \cellcolor{litepurple10}68.23 & \cellcolor{litepurple10}86.94 & \cellcolor{litepurple10}85.61 & \cellcolor{litepurple10}87.85 & \cellcolor{litepurple10}87.70 & \cellcolor{litepurple5}54.96  & \cellcolor{litepurple5}53.29  & \cellcolor{litepurple5}55.49  & \cellcolor{litepurple5}55.14 \\
W+U      & \cellcolor{litepurple5}66.93  & \cellcolor{litepurple5}65.07  & \cellcolor{litepurple5}67.90  & \cellcolor{litepurple5}67.71  & \cellcolor{litepurple5}76.68  & \cellcolor{litepurple5}75.61  & \cellcolor{litepurple5}77.57  & \cellcolor{litepurple5}77.52  & \cellcolor{litepurple5}48.64  & \cellcolor{litepurple5}47.34  & \cellcolor{litepurple5}49.71  & \cellcolor{litepurple5}49.59 \\
W+M      & \cellcolor{litepurple5}65.39  & \cellcolor{litepurple5}64.07  & \cellcolor{litepurple5}67.90  & \cellcolor{litepurple5}67.84  & \cellcolor{litepurple10}87.96 & \cellcolor{litepurple10}86.09 & \cellcolor{litepurple10}88.29 & \cellcolor{litepurple10}87.54 & \cellcolor{litepurple5}53.94  & \cellcolor{litepurple5}52.18  & \cellcolor{litepurple5}54.34  & \cellcolor{litepurple5}54.20 \\
H+W2     & \cellcolor{litepurple10}69.37 & \cellcolor{litepurple10}68.13 & \cellcolor{litepurple10}70.52 & \cellcolor{litepurple10}69.32 & \cellcolor{litepurple10}77.91 & \cellcolor{litepurple10}76.63 & \cellcolor{litepurple10}78.50 & \cellcolor{litepurple10}76.97 & \cellcolor{litepurple5}53.94  & \cellcolor{litepurple5}52.28  & \cellcolor{litepurple5}54.34  & \cellcolor{litepurple5}53.47 \\
H+U      & \cellcolor{litepurple10}69.58 & \cellcolor{litepurple10}68.37 & \cellcolor{litepurple10}70.61 & \cellcolor{litepurple10}69.51 & \cellcolor{litepurple5}73.49  & \cellcolor{litepurple5}72.94  & \cellcolor{litepurple5}74.77  & \cellcolor{litepurple5}74.13  & \cellcolor{litepurple10}57.49 & \cellcolor{litepurple10}56.19 & \cellcolor{litepurple10}58.96 & \cellcolor{litepurple10}58.96 \\
H+M      & \cellcolor{litepurple10}69.37 & \cellcolor{litepurple10}68.96 & \cellcolor{litepurple10}71.52 & \cellcolor{litepurple10}70.22 & \cellcolor{litepurple10}87.91 & \cellcolor{litepurple10}86.34 & \cellcolor{litepurple10}88.69 & \cellcolor{litepurple10}87.33 & \cellcolor{litepurple15}64.94 & \cellcolor{litepurple15}63.28 & \cellcolor{litepurple15}65.32 & \cellcolor{litepurple15}64.74 \\
W2+U     & \cellcolor{litepurple10}67.94 & \cellcolor{litepurple10}66.39 & \cellcolor{litepurple10}68.96 & \cellcolor{litepurple10}67.16 & \cellcolor{litepurple5}75.39  & \cellcolor{litepurple5}74.17  & \cellcolor{litepurple5}76.64  & \cellcolor{litepurple5}76.37  & \cellcolor{litepurple10}57.19 & \cellcolor{litepurple10}56.41 & \cellcolor{litepurple10}58.38 & \cellcolor{litepurple10}58.51 \\
W2+M     & \cellcolor{litepurple5}64.19  & \cellcolor{litepurple5}63.61  & \cellcolor{litepurple5}65.21  & \cellcolor{litepurple5}65.01  & \cellcolor{litepurple10}80.91 & \cellcolor{litepurple10}79.31 & \cellcolor{litepurple10}81.52 & \cellcolor{litepurple10}80.96 & \cellcolor{litepurple15}72.14 & \cellcolor{litepurple15}71.39 & \cellcolor{litepurple15}71.10 & \cellcolor{litepurple15}70.96 \\
U+M      & \cellcolor{litepurple10}68.34 & \cellcolor{litepurple10}67.36 & \cellcolor{litepurple10}69.63 & \cellcolor{litepurple10}67.45 & \cellcolor{litepurple5}73.91  & \cellcolor{litepurple5}72.33  & \cellcolor{litepurple5}74.63  & \cellcolor{litepurple5}73.93  & \cellcolor{litepurple5}49.68  & \cellcolor{litepurple5}48.31  & \cellcolor{litepurple5}50.29  & \cellcolor{litepurple5}49.50 \\
\bottomrule
\end{tabular}
\caption{Evaluation Scores of different PTM combinations; Scores are in \% and average of
five-folds}
\label{fusion}
\end{table*}

\vspace{-0.2cm}

\subsection{Experimental Results}
Table \ref{single} presents the results for individual PTMs with different downstream networks. CNN models generally outperforms SVM and FCN across most models and datasets, with the highest accuracy and F1 scores. FCN performs better than SVM, showing that neural models are better as downstreams with individual PTMs. Among Mamba PTM variants, the base version consistently outperforms the tiny and small versions across all datasets. Its superior performance likely stems from its larger size, enabling better capture of contextual dependencies essential for SER. Also, we can see that PTMs with different downstreams shows variations in results. Such behavior is also reported by previous research \cite{zaiem23b_interspeech}. Among the attention-based PTMs, we observe mixed behavior with some PTMs leading in one dataset and some PTMs in other. This brings out limelight the effect of downstream data distribution on the performance of the downstream task. 
The top performance of MMS in MESD can be traced back to its multilingual pre-training as most of the other PTMs are trained on only English data. However, that's not the case in every scenario, as MMS due to its multilingual pre-training should have good performance in Emo-DB, but it reports one of the lowest performances in comparison to other PTMs. Also, excluding MMS the other attention-based PTMs reported low performance in MESD. In contrast, the some of the attention-based PTMs showed better results than its mamba counterparts in Emo-DB. Overall, there is no clear champion, that mamba or attention-based PTMs are best for SER reinforcing the importance of dataset-specific characteristics in determining the effectiveness of a PTM. \par

\begin{figure}[hbt!]
    \centering
    \subfloat[]{%
        \includegraphics[width=0.20\textwidth]{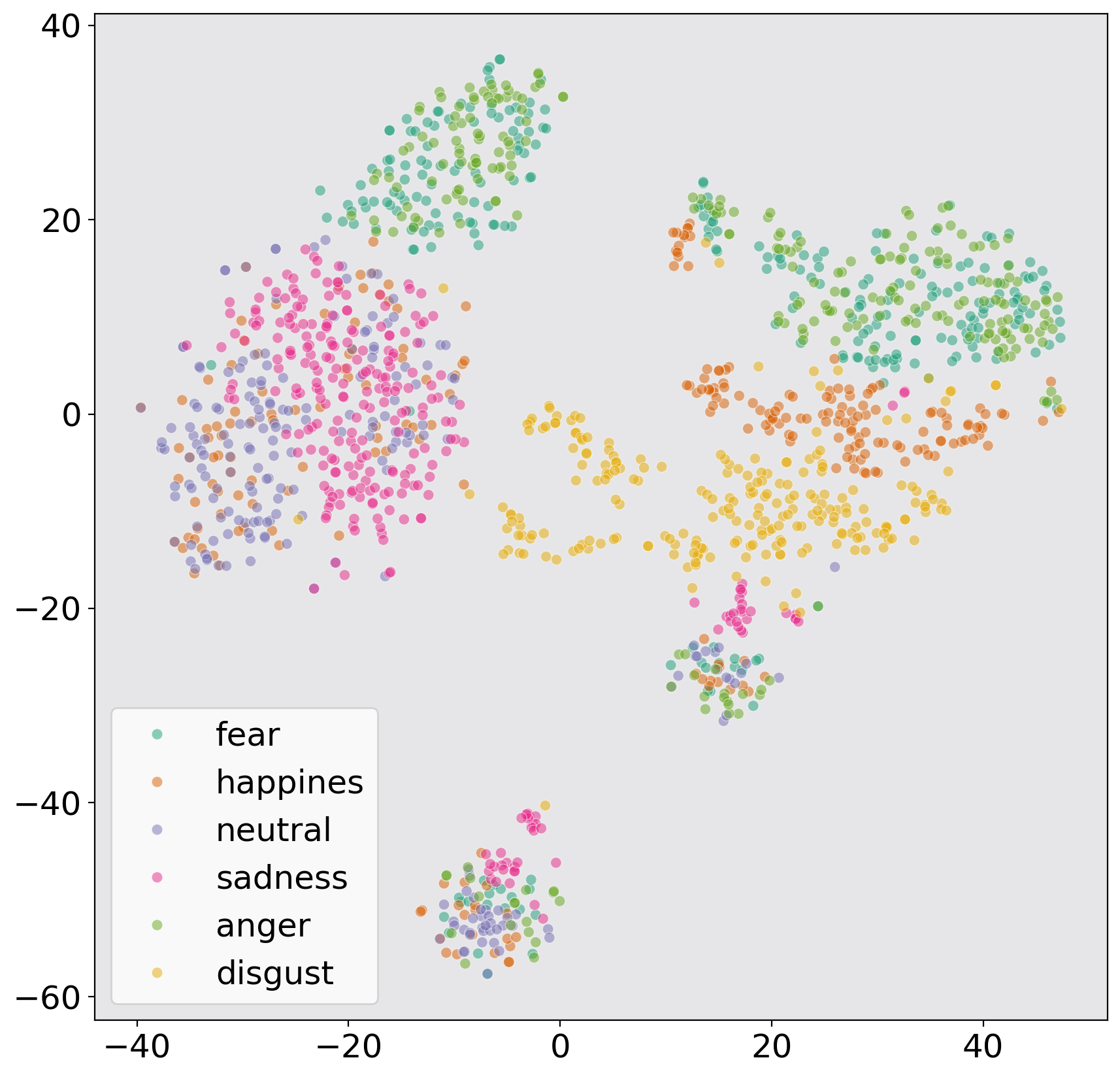}
    }
    \subfloat[]{%
        \includegraphics[width=0.20\textwidth]{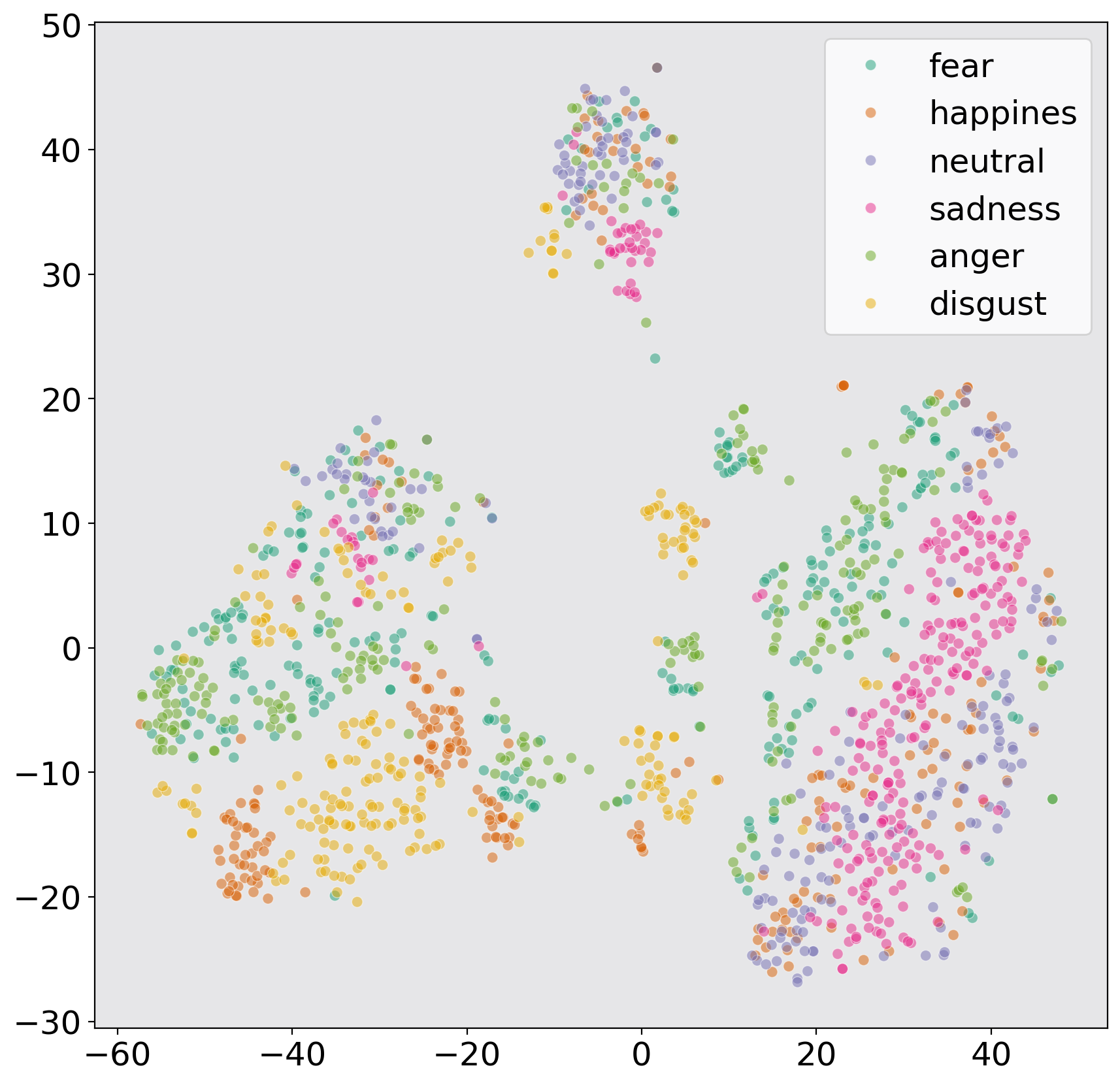}
    }
    \caption{ t-SNE plots for CREMA-D:(a) PARROT with Audio-MAMBA(base) and HuBERT (b) PARROT with Audio-MAMBA (base) and Wav2vec2}
    \label{fig:tsne}
\end{figure}

\vspace{-0.3mm}

\begin{figure}[hbt!]
    \centering
    \subfloat[]{%
        \includegraphics[width=0.20\textwidth]{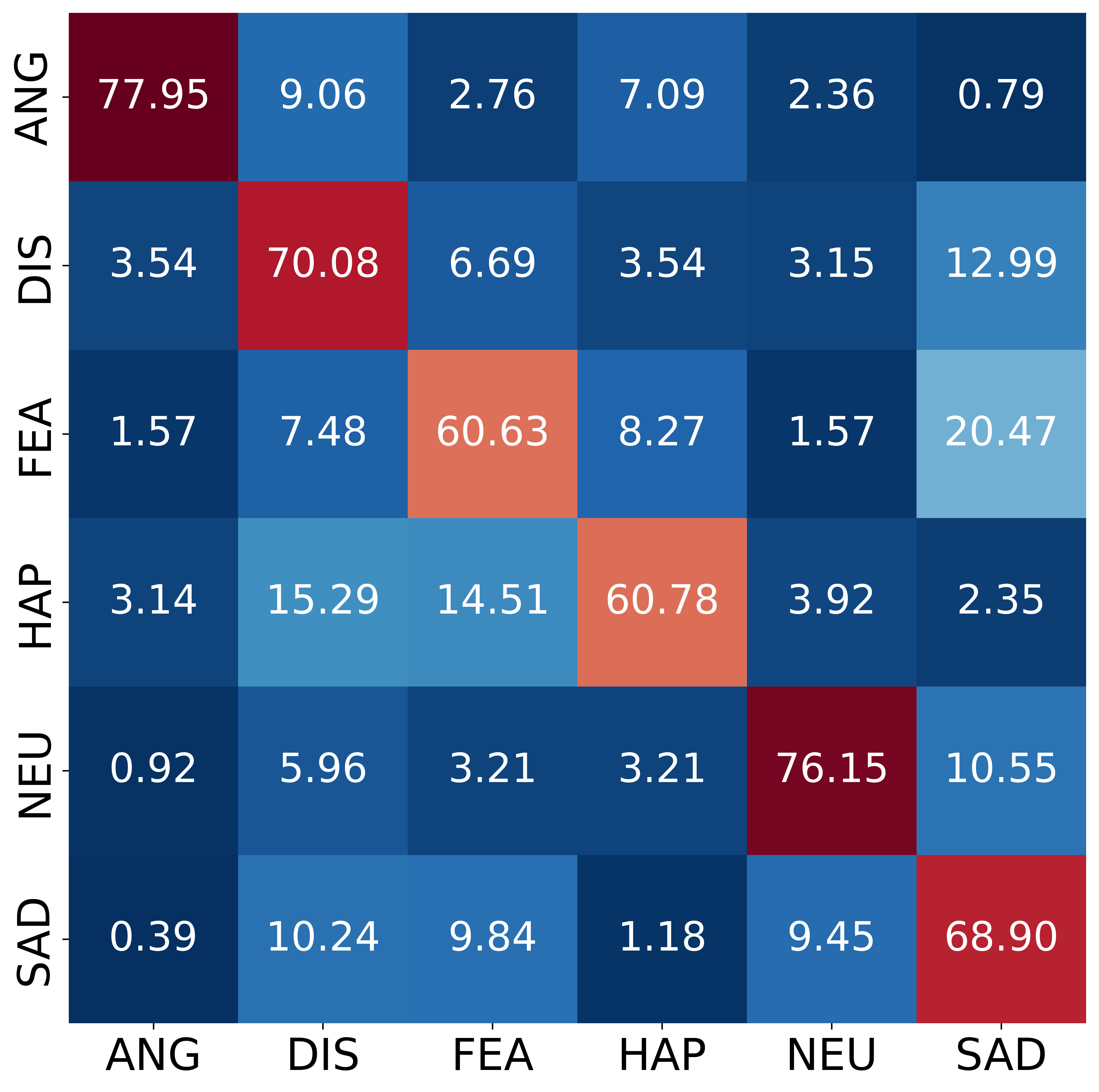}
    }
    \hspace{0.1mm}
    \subfloat[]{%
        \includegraphics[width=0.20\textwidth]{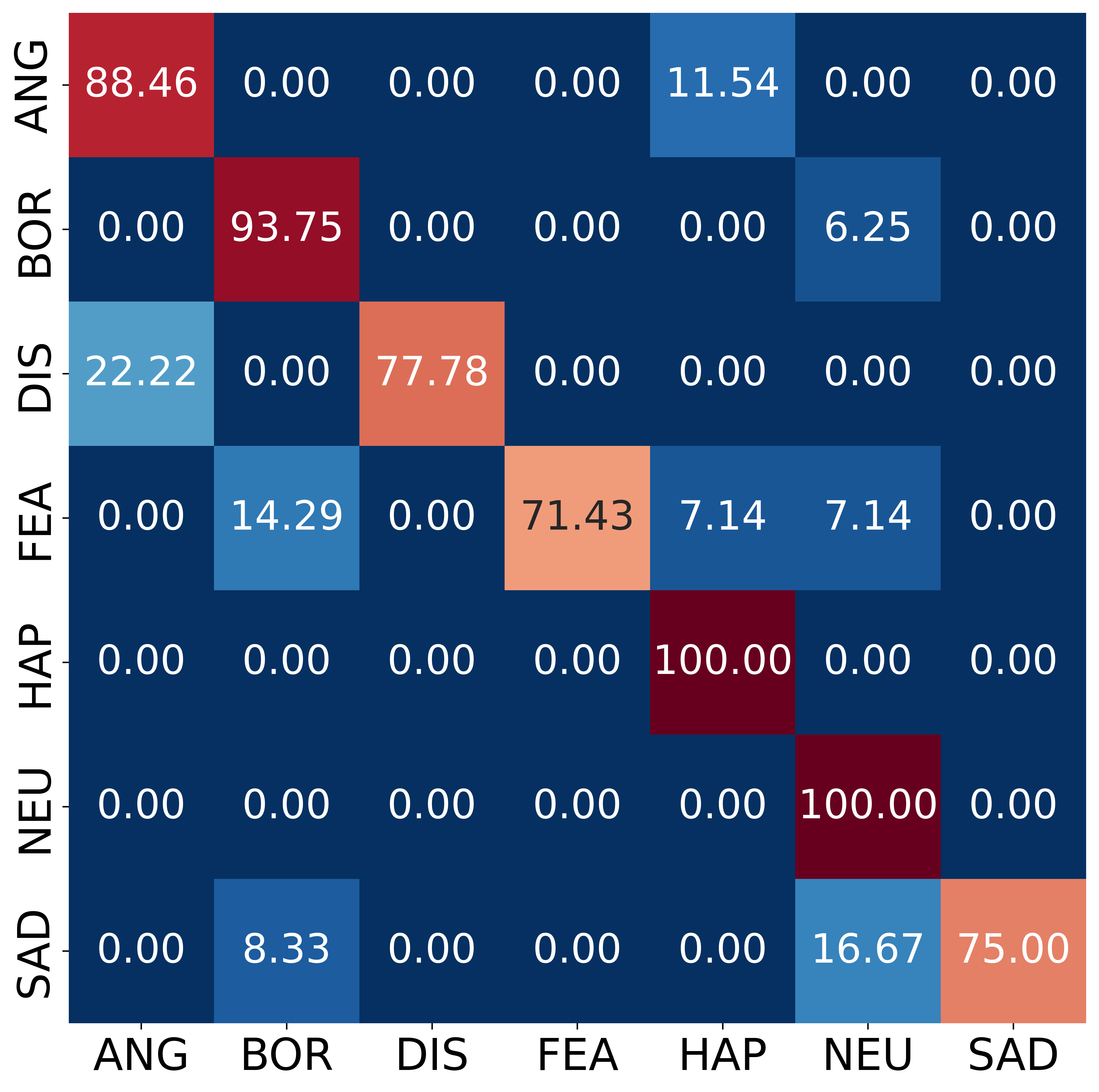}
    }
    \caption{Confusion matrices for PARROT with Audio-MAMBA(base) and HuBERT : (a) CREMA-D (b) EMO-DB; x-axis and y-axis represents predicted and true, respectively}
    \label{fig:cm}
\end{figure}

Table \ref{fusion} shows the results of combinations of different PTMs. We use concatentation-based fusion as the baseline fusion technique. For modeling the concatention-based fusion, we removed the optimal transport and hadamard product parallel branches from \textbf{\texttt{PARROT}} (Figure \ref{fig:parrot_framework}). We kept the rest modeling same and also the training details for fair comparison. Combinations of PTMs through \textbf{\texttt{PARROT}} generally shows better performance in comparison to baseline concatenation-based fusion technique, thus, showing its strength for effective fusion. Further, fusion of different PTMs through \textbf{\texttt{PARROT}} achieves better performance than the individual PTMs across all the datasets. In contrast, fusion of PTMs using concatenation-based fusion overall shows comparable or less performance than individual PTMs except a few specific cases where the fusion of mamba and attention-based PTMs brings strong complementary behavior. This behavior is observed across all the datasets. With \textbf{\texttt{PARROT}}, we observe the emergence of such complementary behavior amongst mamba-and attention-based PTMs in a much better way due to the capability of \textbf{\texttt{PARROT}} to bring out such behavior than baseline concatentation-based fusion technique. For example, fusion of Audio-MAMBA (base) with HuBERT through \textbf{\texttt{PARROT}} reported the topmost performance in CREMA-D and Emo-DB. Also, fusion of Audio-MAMBA (base) with MMS through \textbf{\texttt{PARROT}} reported the best performance in MESD. These results verifies \textit{our hypothesis that heterogeneous fusion of mamba and attention-based PTMs will lead to more improved SER due to the effective emergence of complementary strengths with attention-based PTMs capturing complex global dependencies and mamba-based PTMs excelling in efficient long-range processing.} These results demonstrate that \textbf{\texttt{PARROT}}, by fusing Mamba- and attention-based PTMs, surpasses individual PTMs that previously achieved SOTA performance in SER \cite{yang21c_interspeech, yadav24_interspeech}. Additionally, our approach outperforms most homogeneous PTM fusions and baseline fusion techniques and further establishing its effectiveness in achieving SOTA in SER. We plot the t-SNE plot visualizations of representations from the last penultimate layer in Figure \ref{fig:tsne}. We also plot the confusion matrices of \textbf{\texttt{PARROT}} with fusion of Audio-MAMBA(base) and HuBERT in Figure \ref{fig:cm}.

\vspace{-0.3cm}

\section{Conclusion}

In this study, we explore the heterogeneous fusion of mamba and attention-based SSL PTMs for SER. To this end, we propose, \texttt{\textbf{PARROT}}, a novel framework that synergizes PTMs via parallel branch fusion of Optimal Transport and Hadamard Product. With \texttt{\textbf{PARROT}}, through the fusion of mamba and attention-based PTMs, we report SOTA performance in comparison to individual PTMs, homogeneous fusion of PTMs, and baseline fusion techniques. Our study will act as a reference for future research towards heterogeneous fusion of PTMs for SER.


\bibliographystyle{IEEEtran}
\bibliography{main}

\end{document}